\documentclass{JHEP3}

\title{Hidden conformal symmetry of extreme and non-extreme
Einstein-Maxwell-Dilaton-Axion black holes}

\author{Deyou Chen$^{a}$, Hui Wang, Houwen Wu$^{b}$ and Haitang Yang$^{c}$\\
School of Physical Electronics,\\
University of Electronic Science and Technology of China,\\
Chengdu,  Sichuan 610054, China\\
E-mail: $^{a}$\email{deyouchen@uestc.edu.cn}, $^{b}$\email{iverwu@uestc.edu.cn},
$^{c}$\email{hyanga@uestc.edu.cn}}

\abstract{The hidden conformal symmetry of extreme and
non-extreme Einstein-Maxwell-Dilaton-Axion (EMDA) black holes is
addressed in this paper. For the non-extreme one, employing the
wave equation of massless scalars, the conformal symmetry with
left temperature $T_{L}=\frac{M}{2\pi a}$ and right temperature
$T_{R}=\frac{\sqrt{M^{2}-a^{2}}}{2\pi a}$ in the near region is
found. The conformal symmetry is spontaneously broken due to the
periodicity of the azimuthal angle. The microscopic entropy is
derived by the Cardy formula and is fully in consistence with the
Bekenstein-Hawking area-entropy law. The absorption cross section
in the near region is calculated and exactly equals that in a 2D
CFT. For the extreme case, by redefining the conformal
coordinates, the duality between the solution space and CFT is
studied. The microscopic entropy is found to exactly agree with
the area-entropy law.}

\begin{document}

\section{Introduction}

Holographic duality that reflects the relation of the quantum
gravity theory and quantum field theory plays an important role in
the modern physics \cite{Strominger}. One successful example of
this duality is the AdS/CFT correspondence. The rudiment of this
correspondence was proposed by Maldacena in the late 1990s
\cite{Maldacena1997}. In his work, the equivalence between type
IIB string theory compactified on $AdS_5 \times S^5$ and
four-dimensional supersymmetric Yang-Mills theory was showed. Ever
since then, it has attracted great interest and other examples of
the correspondence have been found. One can refer to
\cite{Maldacena1998} and references therein for reviews.

In \cite{BH1986}, Brown and Henneaux showed that any consistent
theory of quantum gravity in three-dimensional anti-de Sitter
space ($AdS_3$) is holographically dual to a two-dimensional (2D)
conformal field theory (CFT). Inspired by this work \cite{BH1986},
Guica, Hartman, Song and Strominger replaced $AdS_3$ with the
near-horizon extreme Kerr (NHEK) geometry and put forward the
Kerr/CFT correspondence \cite{GHSS}. After specifying the boundary
conditions at the asymptotic infinity of the NHEK, they showed the
asymptotic symmetry group is one copy of the conformal group and
has a central charge $c_L = 12J$. This implies that the quantum
gravity in the NHEK geometry is dual to a 2D CFT. The microscopic
entropy was derived from Cardy formula with the CFT temperatures
and central charges. It turns out that this entropy precisely
agrees with the Bekenstein-Hawking area-entropy law. This result
is a significant step to the evidence of the holographic duality.
The generalizations  to other extreme black holes can be found in
\cite{HHKNT}.

Since the duality of the extreme black holes has been
invesitigated, it is natural to extend this work to near-extreme
and non-extreme black holes. In  \cite{BHSS}, the authors studied
the superradiance of the near extreme black hole and showed the
holographic duality can also exist in this black hole. %In which,
%the scattering amplitudes was reobtained by a dual 2D CFT and in
%which the black hole corresponds to a specific thermal state and
%the scalar field to a specific operator. This work filled in the
%gap that the duality exists in the near extreme black holes.
Very recently, the holographic duality of the non-extreme Kerr
black hole has been explored by Castro, Maloney and Strominger
\cite{CMS}. After introducing the conformal coordinates
($\omega^{+}, \omega^{-}, y$) and the local vector fields, they
found that there is a conformal symmetry in the solution space of
a massless scalar field equation in the near region of the Kerr
black hole. However, this symmetry is local and would be
spontaneously broken due to the existence of the periodicity
($2\pi$) of the azimuthal angle $\phi$. Furthermore, the
absorption cross section in the near region is found to be
precisely the finite-temperature one in a 2D CFT. Moreover, the
entropy, obtained from the Cardy formula with central charges
$c_{L}=c_{R}=12J$, agrees with the Bekenstein-Hawking area-entropy
law. Their results provide more confidence to the duality of
generic Kerr/CFT. Subsequently, this work is extended to other
non-extreme black holes \cite{Krishnan,CS,WL,CL,LLR,Chen:2010zw}.
%
%The hidden conformal symmetry of the general five dimensional
%black hole is studied in \cite{Krishnan} and that of the 5D
%Reissner-Nordstrom black hole, Kerr-Newman and Kerr-Newman-AdS-dS
%black holes, Kaluza-Klein black holes and rotating charged black
%holes were researched in \cite{CS,WL,CL,LLR,Chen:2010zw},
%respectively.
By introducing a new set of conformal coordinates and vector
fields, similar methods are applied to the extreme and near
extreme black holes in \cite{Chen:2010fr}.

The aim of this paper is to extend the work to dilatonic black
holes. We examine the hidden conformal symmetry of the extreme and
non-extreme EMDA black holes. Our result shows that there does
exist a conformal symmetry in the near region. This symmetry is
spontaneously broken under CFT temperatures due to the $2\pi$
periodicity of the azimuthal angle ($\phi$) of the metric. To
clearly show the duality, we calculate the microscopic entropy by
the Cardy formula with central charges $c_{L}=c_{R}=12J$. We find
a full coincidence with the Bekenstein-Hawking area-entropy law.
Meanwhile, the absorption cross section in the near region of the
non-extreme EMDA black hole is derived. This cross section is
exactly equal to the finite-temperature one in a 2D CFT.  We find
that the CFT temperatures are $T_{L}=\frac{M}{2\pi a}$ and
$T_{R}=\frac{\sqrt{M^{2}-a^{2}}}{2\pi a}$ for the non-extreme EMDA
black hole. While the temperatures are $T_{L}=\frac{r_+}{2\pi a}$
and $T_{R}=0$ for the extreme one.

The reminder of this paper is outlined as follows. In sect. 2, we
discuss the hidden conformal symmetry of the non-extreme EMDA black
hole and give the microscopic description. The black hole entropy
and the absorption cross section are derived.  Sect. 3 gives the
hidden symmetry of the extreme EMDA black hole. Sec. 4 is our
discussions and conclusions.

\section{The hidden conformal symmetry of the non-extreme EMDA Black hole}

The EMDA black hole is a solution of the
Einstein-Maxwell-Dilaton-Axion field equations. From the action

\begin{equation}
S=\int d^{4}x\sqrt{-g}\left[R-2g^{\mu\nu}\partial_{\mu}\phi
\partial_{\nu}\phi-\frac{1}{2}e^{4\phi}g^{\mu\nu}\partial_{\mu}\kappa\partial_{\nu}
\kappa-e^{-2\phi}F_{\mu\nu}F^{\mu\nu}-\kappa
F_{\mu\nu}\breve{F}^{\mu\nu}\right],
\end{equation}

\noindent where $\phi$ is the dilaton and $\kappa$ is the axion
scalar field, Garcia, Galtsov and Kechkin \cite{GGK} derived the
axially symmetrical solution in 1995, the EMDA metric,

\begin{eqnarray}
ds^{2} & = &
-\frac{\triangle-a^{2}\sin^{2}\theta}{\Sigma}dt^{2}-\frac{2a\sin^{2}\theta
\left(r^{2}+2br+a^{2}-\triangle\right)}{\Sigma}dtd\varphi\nonumber \\
 &  & +\frac{\Sigma}{\triangle}dr^{2}+\Sigma d\theta^{2}+\frac{\left(r^{2}
 + 2br+a^{2}\right)^{2}-
 \triangle a^{2}\sin^{2}\theta}{\Sigma}\sin^{2}\theta d\varphi^{2},
\end{eqnarray}

\noindent where

\[
\Sigma  =  r^{2}+2br+a^{2}\cos^{2}\theta,
\]
\[
r_\pm = M\pm \sqrt {M^2 - a^2},
\]
\begin{eqnarray}
\Delta  =
r^{2}-2Mr+a^{2}=\left(r-r_{+}\right)\left(r-r_{-}\right).
\end{eqnarray}

\noindent $r_{+}$ and $r_{-}$ are the outer and inner horizons.
The parameters $a$ and $b$ represent the angular momentum and
dilaton constant per unit mass, respectively. The mass $M$ is
related to the ADM mass of the black hole as $M_{A} = M + b$. $J$
is the angular momentum and relates to the black hole mass as $ J
= (M + b) a$. There is a lot of work appeared to study
thermodynamic properties of this black hole. Here we simply give
the thermodynamic parameters, the area $A$, Hawking temperature
$T$ and angular velocity $\Omega$ at the outer horizon,

\begin{eqnarray}
A & = & 4\pi\left(r_{+}^{2}+2br_{+}+a^{2}\right),\nonumber \\
T & = & \frac{r_{+}^{2}-a^{2}}{4\pi
r_{+}\left(r_{+}^{2}+2br_{+}+a^{2}\right)},\nonumber \\
\Omega & =
& \frac{a}{r_{+}^{2}+2br_{+}+a^{2}}.
\end{eqnarray}

We first use Klein-Gordon equation to explore the wave function of
a massless scalar field on the background spacetime of the black
hole,

\begin{equation}
\frac{1}{\sqrt{-g}}\partial_{\mu}\left(\sqrt{-g}g^{\mu\nu}
\partial_{\nu}\Phi\right)=0.
\end{equation}

\noindent Due to the existence of two Killing vectors (
$\partial_{t}$ and $\partial_{\varphi}$) in the EMDA spacetime, we
can carry out  separation of variables as

\begin{equation}
\Phi\left(t,r,\theta,\varphi\right)=e^{-i\omega
t+im\varphi}R\left(r\right)S\left(\theta\right).
\label{wave function} \end{equation}

\noindent Substituting the inverse metric and eqn.(\ref{wave function}) into the
Klein-Gordon equation, one gets two sectors, the angular sector

\begin{equation}
\left[\frac{1}{\sin\theta}\partial_{\theta}\left(\sin\theta
\partial_{\theta}\right)-\frac{m^{2}}
{\sin^{2}\theta}+a^{2}\omega^{2}\cos^{2}\theta\right]
S\left(\theta\right)=-\lambda _{lm}S\left(\theta\right),
\label{angular sector}\end{equation}

\noindent and the radial sector

\begin{eqnarray}
\left[\partial_{r}\triangle\partial_{r}+\frac{\left(2\left(M+
b\right)r_{+}\omega-ma\right)^{2}}{\left(r-r_{+}\right)
\left(r_{+}-r_{-}\right)}-\frac{\left(2\left(M+b\right)r_{-}
\omega-ma\right)^{2}}{\left(r-r_{-}\right)
\left(r_{+}-r_{-}\right)}\right.\nonumber \\
\left.+\left(\triangle-a^{2}+4\left(M+b\right)r\right)
\omega^{2}\right]R\left(r\right) & = & \lambda
_{lm}R\left(r\right), \label{NERadial}
\end{eqnarray}

\noindent where $ \lambda _{lm} $ is the constant of variable
separation. If the term included $\omega^{2}$  vanishes in eqn.(\ref{angular sector}), one obtains the standard Laplacian on the 2-sphere and the
solutions of eqn.(\ref{angular sector}) are spherical harmonics with the
corresponding value $ \lambda _{lm} = l(l+1)$. We follow
\cite{CMS} and only discuss the case of the near region defined by
$r\omega\ll1$. In this case, the terms included $\omega^{2}$
should be neglected and the equations of the angular part as well
as the radial one can be rewritten as

\begin{equation}
\left[\frac{1}{\sin\theta}\partial_{\theta}\left(\sin\theta
\partial_{\theta}\right)-
\frac{m^{2}}{\sin^{2}\theta}\right]S\left(\theta\right)
=-\ell\left(\ell+1\right)S\left(\theta\right),
\end{equation}

\noindent and

\begin{equation}
\left[\partial_{r}\triangle\partial_{r}+\frac{\left(2\left(M+b\right)
r_{+}\omega-ma\right)^{2}}
{\left(r-r_{+}\right)\left(r_{+}-r_{-}\right)}-\frac{\left(2\left(
M+b\right)r_{-}\omega-ma\right)^{2}}
{\left(r-r_{-}\right)\left(r_{+}-r_{-}\right)}\right]R\left(
r\right)=\ell\left(\ell+1\right)R\left(r\right) \label{eq:radial
wave equation}.
\end{equation}

\noindent In the following, we can find that the solution of the
radial part eqn.(\ref{eq:radial wave equation}) is related to $SL\left(2,R\right)$, which
implies the existence of a hidden conformal symmetry. Following
\cite{CMS}, we define the conformal coordinates $(\omega^{\pm}, y
)$ in terms of $(t, r, \phi)$ by

\begin{eqnarray}
\omega^{+} & = & \sqrt{\frac{r-r_{+}}{r-r_{-}}}e^{2 \pi
T_{R}\phi},\nonumber \\
\omega^{-} & = & \sqrt{\frac{r-r_{+}}{r-r_{-}}}e^{2\pi T_{L}\phi -
\frac{t}{2(M+b)}},\nonumber \\ y & = &
\sqrt{\frac{r_{+}-r_{-}}{r-r_{-}}}e^{\pi\left(T_{L}+T_{R}\right)\phi
- \frac{t}{2(M+b)}},
\label{coordinates transformations}\end{eqnarray}

\noindent with $T_{L}=\frac{r_{+}+r_{-}}{4\pi a},
T_{R}=\frac{r_{+}-r_{-}}{4\pi a} $.  The local vector fields are
defined as

\begin{eqnarray}
H_{1} & = & i\partial_{+},\nonumber \\
H_{0} & = & i\left(\omega^{+}\partial_{+}+\frac{1}{2}y
\partial_{y}\right),\nonumber \\ H_{-1} & = &
i\left(\omega^{+2}\partial_{+}+\omega^{+}y\partial_{y}-y^{2}\partial_{-}\right),
\label{vector fields1}\end{eqnarray}

\noindent and

\begin{eqnarray}
\tilde{H}_{1} & = & i\partial_{-},\nonumber \\
\tilde{H}_{0} & = & i\left(\omega^{-}\partial_{-}+\frac{1}{2}
y\partial_{y}\right),\nonumber \\ \tilde{H}_{-1} & = &
i\left(\omega^{-2}\partial_{-}+\omega^{-}y\partial_{y}
-y^{2}\partial_{+}\right),
\label{vector fields2}\end{eqnarray}

\noindent which obey the $SL\left(2,R\right)$ Lie bracket algebra

\begin{equation}
\left[H_{0},H_{\pm1}\right]=\mp
iH_{\pm1},\quad\left[H_{-1},H_{1}\right]=-2iH_{0}.
\label{Lie bracket algebra}\end{equation}
Similar relations are hold for $(\tilde H_0, \tilde H_{\pm})$.
The corresponding quadratic Casimir is

\begin{eqnarray}
\mathcal{H}^{2}=\tilde{\mathcal{H}}^{2} & = &
-H_{0}^{2}+\frac{1}{2}
\left(H_{1}H_{-1}+H_{-1}H_{1}\right)\nonumber \\
 & = & \frac{1}{4}\left(y^{2}\partial_{y}^{2}-y\partial_{y}\right)+y^{2}
 \partial_{+}\partial_{-}.
\label{quadratic Casimir}\end{eqnarray}

\noindent Inserting eqn.(\ref{coordinates transformations}) into eqn.(\ref{quadratic Casimir}) yields

\begin{equation}
\mathcal{H}^{2}=\partial_{r}\triangle\partial_{r}-\frac{\left(a
\partial_{\phi}+2\left(M+b\right)r_{+}
\partial_{t}\right)^{2}}{\left(r-r_{+}\right)\left(r_{+}-r_{-}\right)}+
\frac{\left(a\partial_{\phi}+2
\left(M+b\right)r_{-}\partial_{t}\right)^{2}}{\left(r-r_{-}\right)
\left(r_{+}-r_{-}\right)}.
\label{casimir operator}\end{equation}

\noindent Thus eqn.(\ref{eq:radial wave equation}) can be written as

\begin{equation}
\mathcal{H}^{2}R\left(r\right)=\tilde{\mathcal{H}}^{2}R\left(
r\right)=\ell\left(\ell+1\right)R\left(r\right),
\end{equation}

\noindent which means that the scalar Laplacian can reduce to the
$SL(2,R)$ Casimir and the conformal symmetry exists here. However,
due to the existence of the $2\pi$ periodicity of the azimuth
angle $\phi$ and the local definition of the vector fields, the $
SL(2,R)_L \times SL(2,R)_R $ symmetry is spontaneously broken.

In the following, we give some microscopical interpretation. As
analysis in \cite{CMS}, we find that there exists a relation
between Minkowski $\omega^{\pm}$ and Rindler $t^{\pm}$ coordinates
in non-extreme case,

\begin{equation}
\omega^{\pm}=e^{\pm t^{\pm}},
\end{equation}

\noindent where

\begin{equation}
t^{+}=2\pi T_{R}\phi,\quad t^{-}=\frac{t}{2\left(M+b\right)}-2\pi
T_{L}\phi.
\end{equation}

\noindent This indicates that in the $ SL(2,R)_L\times SL(2,R)_R $
invariant Minkowski vacuum, the observer in Rinder space can
observe the thermal bath of Unruh radiation. Hence, the
non-extreme EMDA black hole is dual to the left and right
temperatures $\left(T_{L},T_{R}\right)$ in a the 2D CFT.

Moreover, one can find the EMDA entropy. It has been known that
the Bekenstein-Hawking area-entropy of black holes can be obtained
by the Cardy formula. However, the central charges $C_L$ and $C_R$
of the non-extreme EMDA black hole are unknown. In Ref.
\cite{CMS,Krishnan,CS,WL,CL,LLR}, the central charges near the
extreme case are adopted and are regarded as those of the
non-extreme case. The reason is that there is a smooth limit from
near-extremal to extremal solution and probably the hidden
conformal symmetry connects smoothly to that of extreme limit.
Therefore we can get the central charges from  the extreme black
holes,

\begin{equation}
c_{L}=c_{R}=12J.
\end{equation}

\noindent Using the Cardy formula with the central charges and
temperatures, we obtain the microcosmic entropy

\begin{equation}
S=\frac{\pi^{2}}{3}\left(c_{L}T_{L}+c_{R}T_{R}\right)=2\pi\left(M+b\right)r_{+},
\end{equation}

\noindent which is the same as the Hawking-Bekenstein area-entropy
of the black hole. We now discuss the solution of the radial
sector. There are many papers appeared to investigate such object
\cite{GBF}. The difference here is that our attention is focused
on the near region $r\omega\ll 1$. After introducing a new
coordinate $ z=\frac{r-r_{-}}{r-r_{+}}$,  eqn.(\ref{NERadial}) becomes

\begin{eqnarray}
 &  & \left(1-z\right)z\partial_{z}^{2}R\left(z\right)+\left(1-z\right)
 \partial_{z}R\left(z\right)\nonumber
 \\
 &  & +\left[\frac{\left(2\left(M+b\right)r_{+}\omega-ma\right)^{2}}{z\left(r_{+}-
 r_{-}\right)^{2}}-\frac{\left(2
 \left(M+b\right)r_{-}\omega-ma\right)^{2}}{\left(r_{+}-r_{-}\right)^{2}}-
 \frac{\ell\left(\ell+1\right)}{1-z}\right]
 R\left(z\right)=0.
 \label{new radial sector}\end{eqnarray}

\noindent Considering the outer boundary, the solution of eqn.(\ref{new radial sector}) takes the form

\begin{eqnarray}
R\simeq Ar^{\ell}+Br^{-\ell-1},
\end{eqnarray}

\noindent where

\begin{equation}
A=\frac{\Gamma\left(1-i2\frac{2\left(M+b\right)r_{+}\omega-ma}{r_{+}
-r_{-}}\right)\Gamma\left(1+2\ell\right)}
{\Gamma\left(1+\ell-i2\left(M+b\right)\omega\right)\Gamma\left(
1+\ell-i\frac{2\left(M+b\right)\left(r_{+}+r_{-}\right)
\omega-2ma}{r_{+}-r_{-}}\right)}.
\end{equation}

\noindent Therefore the absorption cross section is

\begin{eqnarray}
P_{abs}\sim\left|A\right|^{2} & \sim & \sinh\left(\frac{
4\pi\left(M+b\right)r_{+}\omega-2\pi ma}
{r_{+}-r_{-}}\right)\left|\Gamma\left(1+\ell-i2\left(M
+b\right)\omega\right)\right|^{2}\nonumber \\
 &  & \times\left|\Gamma\left(1+\ell-i\frac{2\left(M+
 b\right)\left(r_{+}+r_{-}\right)\omega-2ma}{r_{+}-r_{-}}
 \right)\right|^{2}.
\label{cross section}\end{eqnarray}

\noindent To compare with the finite-temperature absorption cross
section for a 2D CFT, we introduce the conjugate charges $\delta
E_{L}$ and $\delta E_{R}$, which satisfy the relation

\begin{equation}
\delta S=\frac{\delta E_{L}}{T_{L}}+\frac{\delta E_{R}}{T_{R}}.
\label{first law1}
\end{equation}

\noindent The first law of thermodynamics of the black hole tells us

\begin{equation}
T \delta S=\delta M_{A}-\Omega\delta J,
\label{first law}\end{equation}

\noindent where $T, S, M_{A}$ and $\Omega$ are the Hawking
temperature, the area-entropy, the ADM mass and the angular
velocity of the black hole respectively. Combining eqn.(\ref{first law1}) and
(\ref{first law}), one finds

\begin{eqnarray}
\delta E_{L} & = & \frac{\delta M_{A}}{a}\left(M+
b\right)\left(r_{+}+r_{-}\right),\nonumber\\
\delta E_{R} & = &\frac{\delta
M_{A}}{a}\left(M+b\right)\left(r_{+}+r_{-}\right)-\delta J.
\end{eqnarray}

\noindent The left and right moving frequencies are defined as

\begin{eqnarray}
\omega_{L} & \equiv & \delta
E_{L}=\frac{\omega}{a}\left(M+b\right)
\left(r_{+}+r_{-}\right),\nonumber\\
\omega_{R} & \equiv & \delta
E_{R}=\frac{\omega}{a}\left(M+b\right)\left(r_{+}+r_{-}\right)-m,
\label{moving frequencies}\end{eqnarray}

\noindent where $\omega=\delta M_{A}$ and $m=\delta J$ were
introduced. Now applying the conformal weights
$\left(h_{L},h_{R}\right)=\left(\ell,\ell\right)$ and substituting
eqn.(\ref{moving frequencies}) into (\ref{cross section}), one has

\begin{equation}
P_{abs}\sim
T_{L}^{2h_{L}-1}T_{R}^{2h_{R}-1}\sinh\left(\frac{\omega_{L}}
{2T_{L}}+\frac{\omega_{R}}{2T_{R}}
\right)\left|\Gamma\left(h_{L}+i\frac{\omega_{L}}{2\pi
T_{L}}\right)\right|^{2}\left|\Gamma\left(h_{R}+
i\frac{\omega_{R}}{2\pi T_{R}}\right)\right|^{2},
\end{equation}

\noindent which is just the finite-temperature absorption cross
section for a 2D CFT. This gives further evidence to the duality
of the non-extreme EMDA black hole.

\section{The hidden conformal symmetry of the extreme EMDA black hole}

In this section, we address the conformal symmetry of the extreme
EMDA black hole. In the extreme case, the outer horizon and inner
horizon are coincident with each other. This implies that the
conformal coordinate $y$, right temperature and the denominator in
eqs.(\ref{NERadial}), (\ref{eq:radial wave equation}) and (\ref{casimir operator}) are equal to zero. The local vector fields
do not obey the $SL\left(2,R\right)$ algebra in eqn.(\ref{Lie bracket algebra}). Thus
one has to re-construct the conformal coordinates.
%Now we first discuss the radial wave function of the extreme case.
In the extremal case, the radial wave function is different from
that of non-extreme case, eqn.(\ref{eq:radial wave equation}). It
is now

\[
[\partial_{r}\triangle\partial_{r}+\frac{4\left(M+b\right)\omega
\left(\left(M+b\right)r_{+}
\omega-ma\right)}{\left(r-r_{+}\right)}+\frac{\left(2\left(M
+b\right)r_{+}\omega-ma\right)^{2}}{\left(r-r_{+}\right)^{2}}
\]

\begin{equation}
+\left(\triangle-a^{2}+4\left(M+b\right)r+4\left(M+
b\right)^{2}\right)\omega^{2}]R\left(r\right)=
\ell\left(\ell+1\right)R\left(r\right).
\end{equation}

\noindent  We also consider the near region in this section. The
above equation can be rewritten as

\begin{equation}
\left[\partial_{r}\triangle\partial_{r}+
\frac{4\left(M+b\right)\omega\left(\left(M+b\right)
r_{+}\omega-ma\right)}{\left(r-r_{+}\right)}+
\frac{\left(2\left(M+b\right)r_{+}\omega-ma\right)^{2}}{\left(r-
r_{+}\right)^{2}}\right]R\left(r\right)
=\ell\left(\ell+1\right)R\left(r\right).
\end{equation}

\noindent  Due to the coincidence of the outer and inner horizons,
$y$ in the conformal coordinates vanishes. It is necessary to
introduce a new set of conformal coordinates \cite{Chen:2010fr}:

\begin{eqnarray}
\omega^{+} & = & \frac{\gamma_{1}}{2}\left(\frac{1}{a}\phi-
\frac{1}{r-r_{+}}\right),\nonumber\\
\omega^{-} & = & \frac{1}{2}\left(e^{r_{+}\phi/a-\left(M+b\right)
t/2}-\frac{2}{\gamma_{1}}\right),\nonumber\\
y & = & \sqrt{\frac{\gamma_{1}}{2\left(r-r_{+}\right)}}e^{r_{+}
\phi/a-\left(M+b\right)t/2}. \label{ExtremeConCor}
\end{eqnarray}

\noindent The new coordinate transformations have an undetermined
parameter $\gamma_{1}$. This degree of freedom is free to take
arbitrary values (to get the right Casimir operator), based on
particular black holes and they do not affect the physics. The
local vector fields are then defined by eqs.(\ref{vector fields1}) and (\ref{vector fields2}). These
generators again satisfy the $SL(2,R)$ algebra, eqs.(\ref{Lie bracket algebra}) and
(\ref{quadratic Casimir}). Inserting the above equations into (\ref{quadratic Casimir}), one derives the
Casimir

\begin{equation}
\mathcal{H}^{2}=\partial_{r}\triangle\partial_{r}-\frac{4\left(
M+b\right)\partial_{t}\left(2\left(M+b\right)
r_{+}\partial_{t}+a\partial_{\phi}\right)}{
\left(r-r_{+}\right)}-\frac{\left(2\left(M+b\right)r_{+}
\partial_{t}+
a\partial_{\phi}\right)^{2}}{\left(r-r_{+}\right)^{2}}.
\end{equation}

\noindent Hence the  radial wave function can be written as

\begin{equation}
\mathcal{H}^{2}R\left(r\right)=\tilde{\mathcal{H}}^{2}R\left(r\right)
=\ell\left(\ell+1\right)R\left(r\right),
\end{equation}

\noindent Thus, the conformal symmetry exists in the extreme EMDA
black hole. Similarly, this symmetry is spontaneously broken due
to the the existence of the periodicity of the azimuth angle
($\phi - \phi+2\pi$).
%Now the CFT left temperature is $T_L =
%\frac{r_+}{2\pi a}$ and the right temperature is zero.
In the extreme case, as indicated by the second equation in eqn.(\ref{ExtremeConCor}), there only exists one relation
$\omega^{-}=e^{-t^{-}}$  between Minkowski $\omega^{-}$ and
Rindler $t^{-}$, where $t^{-}=\left(M+b\right)t/2-2\pi T_{L}\phi$,
and $T_{L}=r_{+}/2\pi a$. It implies that an observer in the
Rinder space can observe the thermal bath of Unruh radiation and
the extremal EMDA black only dual to the left temperature
%$\left(T_{L},0\right)$
in the 2D CFT. Therefore, the CFT left temperature is $T_L =
\frac{r_+}{2\pi a}$ and the right temperature vanishes.

The asymptotic symmetry group (ASG) plays an important role in the
studies of Kerr/CFT
correspondence. %As  mentioned in \cite{GHSS},
Every consistent set of boundary conditions specifies an
associated ASG for the near horizon extreme black holes. The
Virasoro algebra is obtained from the generators of the ASG. The
corresponding central charge is related to the angular momentum as
$ c_L = 12\pi J$. Using the Cardy formula with the central charges
derived, the microscopic entropy is

\begin{equation}
S=\frac{\pi^{2}}{3}c_{L}T_{L}=2\pi\left(M+b\right)r_{+},
\end{equation}

\noindent which is agreement with the Bekenstein-Hawking
area-entropy.

\section{Conclusions}

In conclusion, we have discussed the hidden conformal symmetry in
the solution space of the massless scalar for the extreme and
non-extreme EMDA black holes. We found that there is a
spontaneously symmetry breaking due to the periodicity $(2\pi)$ of
the azimuth angle $ \phi $. The Bekenstein-Hawking area-entropies
of the black holes were recovered by the Cardy formula with the
central charges. We also found that the CFT temperatures
$T_{L}=\frac{M}{2\pi a}$ and $T_{R}=\frac{\sqrt{M^{2}-a^{2}}}{2\pi
a}$ for the non-extreme EMDA black hole and that
$T_{L}=\frac{r_+}{2\pi a}$ and $T_{R}= 0$ for the extreme one.
Finally, we investigated the absorption cross section of the
non-extreme EMDA black hole in the near region. This cross section
is equal to the finite-temperature one in a 2D CFT. Our results
provide evidence for the duality between the near region of the
EMDA background spacetime and the two-dimensional CFT.

The investigation of the hidden conformal symmetry in this paper
is limited to the scalar fields. It is natural to expect that the
symmetry also exists in the solution space of higher spin fields.
Thus it is of interest to conduct research on  the spinor fields
or vector fields in the future work.

\acknowledgments
This work is supported in part by Fundamental Research Funds for
the Central Universities (Grant No. ZYGX2009X008),  NSFC (Grant
No.10705008) and NCET.

\end{document}